\documentclass[12pt]{article}
\usepackage{amsfonts}
\usepackage{amssymb}
\usepackage{amsthm}
\usepackage{hyperref}
\usepackage{amssymb,amsmath,amsthm}
\usepackage{enumerate}
\usepackage{graphicx}

\newcommand{\sss}{\setcounter{equation}{0}}
\newtheorem{theorem}{THEOREM}[section]

\newtheorem{definition}[theorem]{DEFINITION}

\newcommand{\ere}{ {\mathbb R}}
\def\cl{{\mathbb R}_0^3}

\newcommand{\CE}{{\mathbb C}}
\def\beq{\begin{equation}}
\def\ene{\end{equation}}
\def \ds {\displaystyle}
\newcommand{\bull}{\hfill $\Box$}

\def\qed{\ifhmode\unskip\nobreak\fi\ifmmode\ifinner
\else\hskip5pt\fi\fi\hbox{\hskip5pt\vrule width4pt height6pt
depth1.5pt\hskip1pt}}

\def\var{\varepsilon}

\def\e{\mathbf E}
\def \b{\mathbf B}
\def\d{\mathbf D}
\def\h{\mathbf H}
\def\x{\mathbf x}
\def\y{\mathbf y}
\def\c{\mathbf c}
\def\H{\mathcal H}
\def\C{\mathbf C}

\def\grad{\,\hbox{grad}_{\theta \varphi}\,}
\begin{document}
\baselineskip=20 pt
\parskip 6 pt

\title{The Boundary Conditions for Point Transformed
Electromagnetic Invisibility  Cloaks
\thanks{ PACS classification scheme 2006: 41.20.Jb, 02.30.Tb,02.30.Zz, 02.60.Lj. AMS 2000 classification
35L45, 35L50, 35L80, 35P25, 35Q60, 78A25, 78A45.} \thanks{ Research partially
supported by  CONACYT under Project P42553­F.}}
 \author{  Ricardo Weder\thanks{ Fellow Sistema Nacional de Investigadores.}
\\ Departamento de M\'etodos
 Matem\'aticos  y Num\'ericos. \\Instituto de Investigaciones en Matem\'aticas Aplicadas y en Sistemas.\\
 Universidad Nacional Aut\'onoma de
M\'exico.\\ Apartado Postal 20-726, M\'exico DF 01000\\
 weder@servidor.unam.mx}

\date{}
\maketitle
\begin{center}
\begin{minipage}{165mm}
\centerline{{\bf Abstract}}
\bigskip
In this paper we study point transformed electromagnetic
invisibility cloaks in transformation media that are obtained by
transformation from general anisotropic media. We assume that there
are several point transformed electromagnetic cloaks located in different points in space. Our results
apply in particular to the first order invisibility cloaks
introduced by Pendry et al. and to the high order invisibility
cloaks introduced by  Hendi et al. and by Cai et al.. We identify
the appropriate {\it cloaking boundary conditions} that the
solutions of Maxwell equations have to satisfy at the outside,
$\partial K_+$, and at the inside, $\partial K_-$, of the boundary
of the cloaked object $K$ in the case where the permittivity and the permeability are bounded below and above in $K$.
Namely, that the tangential components of
the electric and the magnetic fields have to vanish
 at $\partial K_+$ - what is always true-  and that the normal components of the curl of the electric and the magnetic fields have to vanish
 at $\partial K_-$.
These results are proven  requiring that energy be conserved. In the
case of one  spherical cloak with  a spherically  stratified $K$ and
a radial current at  $\partial K$ we verify by an explicit
calculation that our {\it cloaking boundary conditions} are
satisfied and that cloaking of active devices holds, even if the
current is at the boundary of the cloaked object. As we prove our
results for media that are obtained by transformation from general
anisotropic media, our results apply to the cloaking of objects with
 passive and active devices contained in general anisotropic media, in
particular to objects with  passive and active devices contained
inside general crystals.
\end{minipage}
\begin{minipage}{165mm}
Our results suggest a method to enhance
cloaking in the approximate transformation media that are used in
practice. Namely, to coat the boundary of the cloaked object (the
inner boundary of the cloak) by a material that imposes the boundary
conditions above. As these boundary conditions have to be satisfied
for  exact transformation media, adding a lining that enforces them
in the case of approximate transformation media will  improve the
performance of approximate cloaks.
\end{minipage}
\end{center}


\section{Introduction}\sss

In this paper we study point transformed electromagnetic
invisibility cloaks in transformation media that are obtained by
transformation from general anisotropic media. We assume that there
are several cloaks located in different points in space. Our results
apply in particular to the first order invisibility cloaks
introduced by \cite{pss1} and to the high order invisibility cloaks
introduced by \cite{hen, ho}.

In \cite{we1}, \cite{we2} we gave a rigorous proof -based in energy conservation- of cloaking of passive and active devices for our general class
of invisibility  cloaks.
The cloaked object, $K$, completely decouples from the exterior. Actually, the cloaking
outside is independent of what is inside $K$. The electromagnetic waves inside
$K$ can not leave $K$ and vice versa, the electromagnetic waves outside
 can not go inside.
Furthermore, we  identified  the appropriate {\it cloaking boundary conditions} when cloaking is
formulated as a boundary value problem. We proved that the tangential components of the electric and the magnetic
fields have to vanish at the outside of the boundary of the cloaked object, $\partial K_+$.  This boundary condition is
self-adjoint in our case because the permittivity and the permeability are degenerate at $\partial K_+$. We also proved that the boundary condition at
the inside of the boundary of the cloaked object, $\partial K_-$, can be any self-adjoint boundary condition for the Maxwell generator in $K$ .
This is true in the general case where the permittivity and the permeability are allowed to be degenerate at $\partial K_-$ . In this general situation
the
particular boundary condition that nature will take on $\partial K_-$ will depend on the behaviour of the permittivity and the permeability near
$\partial K_-$.
We proved our results both in the time and in the frequency domains, and as we consider media obtained by transformation from general anisotropic media,
our results  apply, in particular, to objects inside general crystals.

In this paper we address the problem of determining the {\it
cloaking boundary conditions} at $\partial K_-$ in the case where
the permittivity and the permeability inside $K$ are bounded  and
have a positive lower bound. This corresponds to the situation where
we have a standard object $K$ -that could be anisotropic and
inhomogeneous, but whose permittivity and permeability  are neither
singular nor degenerate- that is coated by a transformation medium
that is degenerate at $\partial K_+$. We also allow for active
devices in $K$. This is perhaps the more important case in the
applications. We prove that in this case the  {\it cloaking boundary
conditions} at $\partial K_-$ are that the normal components of the
curl of the electric and the magnetic fields  vanish.

 Since we have identified the {\it cloaking boundary conditions} at $\partial K_\pm$ we have now a complete formulation of cloaking as a
 boundary value
 problem in this important case. For  the exact transformation media that we consider in this paper these  boundary conditions are satisfied
  because they
follow from energy conservation and there is no need to add any
lining to impose them. In other words, they are the conditions taken
by nature, as they are imposed by energy conservation. However, our
results suggest a method to enhance cloaking in the approximate
transformation media that are used in practice. Namely, to coat the
boundary of the cloaked object (the inner boundary of the cloak) by
a material that imposes the boundary conditions above. As these
boundary conditions have to be satisfied for  exact transformation
media, adding a lining that enforces them in the case of approximate
transformation media will  improve the performance of approximate
cloaks.

It is, of course, a well known fact  in electromagnetic theory -and
in wave propagation in general- that in any interphase between two
different media there has to be a boundary condition. This obviously
applies to the interphase between the cloaked object and the coating
metamaterial, i.e., $\partial K$. So, the real question is not if
there has to be  boundary conditions at $
\partial K_{\pm}$, but rather what are the appropriate boundary
conditions. We address this question in this paper as well as in
\cite{we1,we2}. The reason why this is a delicate problem that
requires a careful analysis is that for point transformed
electromagnetic cloaks the permittivity and the permeability are
degenerate at $\partial K$ and in consequence the standard rules
that are used in the  non-degenerate case do not apply. In fact, the
solutions to Maxwell equations are, in general, discontinuous at
$\partial K$.

The interesting paper \cite{yyrq} considers point and line
transformed electromagnetic cloaks under general coordinate
transformations. Among other problems, they compute the fields
outside the cloaked object from the fields in the original
electromagnetic space, using the transformation formulae between
them, what avoids doing tedious calculations in the transformed
space. This method was previously used in \cite{we1,we2} for
spherical and cylindrical cloaks. In this way  it is proven in
\cite{yyrq}  that for general point transformed cloaks the
tangential components of the electric and the magnetic fields vanish
at the outside of the boundary of the cloaked object and also that
for general line transformed cloaks the tangential components of the
electric and magnetic fields that are orthogonal to the axis of the
cloak  vanish at the outside of the boundary of the cloaked object.
This generalizes the results previously proved in \cite{we1,we2} in
the case of spherical and cylindrical cloaks, using the same method.

The paper \cite{gklu} considers cloaking  in terms of the Cauchy
data, in the context of the Dirichlet to Neumann operator. Among
other problems, they study cloaking of passive and active devices
for one spherical electromagnetic cloak. They postulate a class of
weak solutions in distribution sense across $\partial K$ (see
Definition 4 of \cite{gklu}). They study the case when the
permittivity and permeability are bounded below and above inside $K$
( what they call the single coating)  in Theorem 5, where, among
other results, they prove that the tangential components of the
electric and magnetic fields of their solutions have to vanish at
$\partial K_-$. They conclude that their solutions do not exists for
generic currents inside $K$, and that cloaking holds for passive
devices but that it fails for active devices with generic currents
inside $K$. To deal with this issue they propose to add a perfect
electric conducting lining to $K$ -what makes it to appear as
passive- or to introduce a different construction that they call the
double coating.

In this paper, as well as in \cite{we1,we2}, we proceed in a
completely different way. Instead of postulating {\it a priori} a
particular class of weak solutions in distributions sense across
$\partial K$ we first characterize all possible ways to define
solutions that are compatible with energy conservation. They
correspond to all self-adjoint extensions of the Maxwell generator.
Note that each self-adjoint extension can be understood in terms of
boundary conditions at $\partial K_\pm$. We proved in \cite{we1,we2}
that all self-adjoint extensions are the direct sum of some
self-adjoint extension inside $K$ with a fixed self-adjoint
extension outside $K$. This implies that the solutions inside and
outside of $K$ are completely decoupled from each other and that, in
general, they are  discontinuous at $\partial K$. Another consequence
is that in the case where the permittivity and the permeability are
bounded below and above inside $K$ weak solutions in distribution
sense across $\partial K$, and more generally solutions with
transmission conditions that link the inside and the outside of $K$,
are not self-adjoint, i.e., they are not allowed by energy
conservation.  Note that in this case,  as the permittivity and the
permeability are bounded below and above inside $K$, requiring that
 the tangential components  of both, the electric and the magnetic field vanish at $\partial K_-$ is not
 a self-adjoint boundary condition. We  only are allowed to require that one of them vanishes.
 However, requiring that both vanish at $\partial K_+$ is
 a self-adjoint boundary condition because the  permittivity and the
permeability are degenerate at $\partial K_+$.

We also proved in \cite{we1,we2} that cloaking of passive and active
devices always holds for all possible ways to define solutions that
satisfy energy conservation, i.e., with self-adjoint boundary
conditions.

There are many papers that discuss line transformed, or cylindrical,
cloaks. See for example \cite{pss1, ho,  we1,yyrq, gklu, ccks, cpss,
smjcpss, zcwk, sps, gklu2, rynq, zcwrk}. In \cite{gklu2}
boundary conditions are considered to enhance cloaking for a
cylindrical cloak. Note that the results for line transformed -or
cylindrical- cloaks are quite different from the ones for point
transformed cloaks studied in this paper.

As it is often the case in the papers on electromagnetic
invisibility cloaks, we make the assumption that the media are not
dispersive. This is a widely used idealization. As is well known,
metamaterials are dispersive, and, furthermore, when the
permittivity and the permeability have eigenvalues less than one,
dispersion comes into play in order that the group velocity does not
exceeds the speed of light. This idealization means that we have to
take a narrow enough range of frequencies in order that we can
analyze the cloaking  effect without taking dispersion into account.
In practice this means that cloaking will only be approximate. Note,
moreover, that the results in  this paper, as well as in
\cite{we1,we2}, are proven for cloaks  in exact
transformation media.

The paper is organized as follows. In Section 2 we prove that our {\it cloaking boundary conditions}  at $\partial K_-$ are satisfied. In Section 3
 we illustrate our method by considering   one spherical cloak   with an active device given by a
radial electric current  at the boundary of $K$. The case of a
magnetic current at the boundary of $K$ follows in the same way. We
assume that $K$ is isotropic and spherically stratified. We verify
in this particular case,  by an explicit computation, that our {\it
cloaking boundary conditions} are satisfied and that cloaking of
active devices holds, even if the current is at the boundary of the
cloaked object, as we have proven in Section 2 in the general case
where there is no explicit solution. We end the paper with
conclusions.

 For other results on invisibility cloaks see references \cite{cwzk,le,shd} as well as  the references quoted there and in \cite{we1,we2}.

\section{The Boundary Conditions}
\sss
Let us consider Maxwell equations in $\ere^3$, in the time domain,

\begin{eqnarray}
\nabla \times \e &=& -\frac{\partial}{\partial t}\b, \,\, \nabla \times \h \,=\,\frac{\partial }{\partial t}
 \d,
\label{2.1} \\\nonumber\\
 \nabla \cdot \b&=&0, \nabla \cdot \d\,=
 \,0,
\label{2.2}
 \end{eqnarray}
and in the frequency domain, assuming a periodic time dependence of
$\e,\h$  given by \linebreak $e^{iw t}$, with $\omega$ the
frequency,
\begin{eqnarray}
\nabla \times \e &=& - i\omega \b, \,\, \nabla \times \h
\,=\,i\omega \d,\,\, \omega \neq 0,
 \label{2.3}\\
 \nabla \cdot \b&=&0, \nabla \cdot \d\,=
 \,0,
\label{2.4}
 \end{eqnarray}
where we have suppressed the factor $e^{i\omega t}$ in both sides.  In this paper we take  the time factor $e^{i\omega t}$ to use the  convention
 of \cite{vb, vhu}.  Note that (\ref{2.4})
follows from (\ref{2.3}).

We briefly recall some notations and definitions from \cite{we2}.

 Let us first consider the case where there is only one cloak located at $\x =0$.
See Figure \ref{fig1}. We designate the Cartesian coordinates of $\x$ by $x^\lambda, \lambda=1,2,3$. To define
 the transformation media we introduce another copy of $\ere^3$, denoted by $\cl$. The points in $\cl$ are
 denoted by $\y$ with coordinates
$y^\lambda, \lambda=1,2,3$. We designate, $\hat{\x}:= \x/|\x|, \hat{\y}:= \y/|\y|$.
 Consider the following transformation from
$\cl\setminus \{0\}$ to $\ere^3$,

\beq
\x=\x(\y)=f(\y):= g(|\y|) \hat{\y}.
\label{2.5}
\ene
In spherical coordinates this transformation changes the radial coordinate but leaves the angular
coordinates constant, i.e., $|\x|= g(|\y|), \hat{\x}=\hat{\y}$. Given $ 0 <a < b$  we wish that this
transformation sends the punctuated ball $ 0 < |\y| \leq b$ onto the concentric shell $ a < |\x| \leq b$,
that it is the
identity for $|\y| \geq b$ and that it is one-to-one. Then, we assume that  $g$ satisfies the following
conditions.

\begin{definition}\label{def-2.1}
For any positive numbers $a, b $ with $ 0 <a < b$,  we say the $g$ is a cloaking function  in
$[0,b]$ if  $g(\rho)$ is twice continuously
differentiable on $[0,b],  g(0)=a, g(b )=b$,  and
$g'(\rho):=\frac{d}{d\rho}g(\rho)> 0, \rho \in [0,b]$.
\end{definition}

We define,

\beq \begin{array}{c}
\x=\x(\y)=f(\y):= g(|\y|) \hat{\y},\, \hbox{\rm for}\,0< |\y|\leq b, \\
\x=\x(\y):= \y,\, \hbox{\rm for}\, |\y| \geq b.
\end{array}
\label{2.6}
\ene

With these conditions (\ref{2.6}) is a bijection from $\cl \setminus\{0\}$
onto $\ere^3 \setminus B_a(0)$, where we denote

\beq
B_r(\x_0):= \left\{ \x \in \ere^3: |\ x- \x_0| \leq r \right\}.
\label{2.7}
\ene
Moreover, it blows up the point $0$ onto the sphere
$|\x|=a$. It sends the punctuated ball   $ 0 < |\y| \leq b$ onto the concentric shell $ a <|\x | \leq b$
and  it is the identity for $ |\y| \geq b$.
 It is twice continuously  differentiable away from the sphere $|\y|=b$, where it can have
 discontinuities in the derivatives depending on the values of the derivatives of $g$ at $b$.

 In \cite{ho}  the quadratic case
$$
g(\rho)= \left[ 1-\frac{a}{b}+ p(\rho -b)\right] \rho +a
$$
 with $ p \in \ere$  was discussed in connection with a cylindrical cloak in an approximate
 transformation medium.
 In \cite{pss1} the first order  case $ g(\rho)= \frac{b-a}{b}\rho +a$  was considered. First order transformations were previously used in
 \cite{glu1, glu2} in the context of Calder\'on's inverse conductivity problem.

The closed ball $K:=\{ \x \in \ere^3: |\x|\leq a\}$ is the region that we wish to conceal, and we call it the
cloaked object. The  spherical shell $a < |\x| \leq b$ is the cloaking layer. The union of the
cloaked object  and the cloaking layer is the spherical cloak. The domain  $ |\x| > b$ is the exterior of the spherical cloak.

We now put a finite number of  spherical cloaks in different points in space
in such a way that they do not intersect. See Figure \ref{fig2}.
Let us take  as
centers of the cloaks points  $ \mathbf{c}_j \in \ere^3, j=1,2,\cdots N$ where $N$ is the number of cloaks
 and
${\mathbf c}_j \neq {\mathbf c}_l, j \neq l,  1\leq j,l  \leq N$. We take $ 0 < a_j < b_j,
 \,\hbox{\rm and cloaking functions}\, g_j\, $  that satisfy the conditions of Definition
 \ref{def-2.1} for $a_j,b_j, j=1,2,3\cdots N$,
and we define the following transformation from $ \cl\setminus \{\c_1,\c_2,\cdots , \c_N\}$ to $\ere^3$.

\beq
\begin{array}{c}
 \x=\x(\y) = f(\y):= \c_j + g_j(|\y-\c_j|) \,\, \widehat{\y-\c_j}, \y \in B_{b_j }({\mathbf c}_j),
\, j=1,2, \cdots, N, \\\\
 \x=\x(\y) = f(\y):=\y , \y \in  \cl \setminus \cup_{j=1}^N B_{b_j }({\mathbf c}_j),
\end{array}
\label{2.8}
\ene
where   $B_{b_j }({\mathbf c}_j)$ are balls in $\cl$.

The cloaked objects that we wish to conceal are  given by,

\beq
K_j:=\left\{ \x  \in \ere^3 : |\x-\c_j |\leq   a_j \right\}, j=1,2,\cdots,N.
\label{2.9}
\ene
The  spherical shells  $a_j < |\x- \c_j| \leq b_j, j=1,2,\cdots,N$
 are  the cloaking layers.  The spherical cloaks are  the balls $B_{b_j}(\c_j)$ in $\ere^3$. We denote by
$K$  the union of all the cloaked objects,
\beq
K:= \cup_{j=1}^N K_j.
\label{2.10}
\ene
The domain
\beq
\ere^3 \setminus \cup_{j=1}^N B_{b_j}(\c_j)
\label{2.11}
\ene
is the exterior of the all the spherical cloaks.
We assume that   the spherical cloaks are at a positive distance of each other,

$$
\hbox{min distance}\left( B_{b_j}(\c_j),  B_{b_l}(\c_l) \right) >0, j\neq l, j,l =1,2,\cdots, N.
$$
Denote
$$
\Omega_0:= \cl \setminus \{\c_1, \c_2, \cdots,\c_N\},\,\, \Omega:= \ere^3 \setminus K.
$$

 Then, (\ref{2.8}) is a    bijection from $\Omega_0$ onto
$\Omega$,   and  for $j=1,2, \cdots,N$ it blows up the point $\c_j$ onto the sphere
$|\x- \c_j|=a_j$. It sends the punctuated ball $ 0 < |\y-\c_j| \leq b_j$ onto the
shell $ a_j <|\x -\c_j| \leq b_j$
and  it is the identity for $ \y \in  \cl \setminus \hbox{\rm interior}\left(\cup_{j=1}^N B_{b_j }
(\c_j)\right)$. It is twice continuously differentiable away from the spheres $|\y-\c_j|=b_j$, where it
can have
 discontinuities in the derivatives depending on the values of the derivatives of $g_j$ at $b_j$.

 The elements of the Jacobian matrix are denoted by
$ A^\lambda_{\lambda'}$,
\beq
A^\lambda_{\lambda'}:= \frac{\partial x^\lambda}{\partial y^{\lambda'}}.
\label{2.12}
\ene
$A^\lambda_{\lambda'}\in
 \C^1\left(\Omega_0 \setminus \cup_{j=1}^N \partial B_{b_j}(\c_j)\right)$, and that it can have discontinuities
 on $\cup_{j=1}^N \partial B_{b_j}(\c_j)$ depending on the derivatives of $g_j$ at $b_j$.
 We designate by $A^{\lambda'}_\lambda$ the elements
of the Jacobian of the inverse bijection,
$\y=\y(\x)=f^{-1}(\x)$,

\beq
A^{\lambda'}_{\lambda}:= \frac{\partial y^{\lambda'}}{\partial x^{\lambda}}.
\label{2.13}
\ene
$A^{\lambda '}_{\lambda}\in
 \C^1\left(\Omega \setminus \cup_{j=1}^N \partial B_{b_j}(\c_j)\right)$, and it can have discontinuities
 on $\cup_{j=1}^N \partial B_{b_j}(\c_j)$ depending on the derivatives of $g_j$ at $b_j$.

It follows from  (\ref{2.8}) that the transformation matrix (\ref{2.12}) is given by,

\beq
\begin{array}{c}
 A^\lambda_{\lambda'}= \frac{g_j(|\y-\c_j|)}{|\y-\c_j|} \delta^\lambda_{\lambda'}+
 \left( \frac{g_j'(|\y-\c_j|)}{|\y-\c_j|^2}-\frac{g_j(|\y-\c_j|)}{ |\y-\c_j|^3}\right)
 (\y-\c_j)^\lambda (\y-\c_j)^{\lambda'},\\\\
 \y \in B_{b_j }({\mathbf c}_j),
1 \leq j \leq N, \\\\
  A^\lambda_{\lambda'} = \delta^\lambda_{\lambda'}\ , \y \in  \cl
  \setminus \cup_{j=1}^N B_{b_j }({\mathbf c}_j).
\end{array}
\label{2.14}
\ene
The determinant is equal to,
\beq
\begin{array}{c}
\Delta(\y)= g_j'(|\y-\c_j|) \left( \frac{g_j(|\y-\c_j|)}{|\y-\c_j|}\right)^2,
 \y \in B_{b_j }({\mathbf c}_j),
1 \leq j \leq N, \\\\
  \Delta(\y)=1 , \y \in  \cl
  \setminus \cup_{j=1}^N B_{b_j }({\mathbf c}_j).
\end{array}
\label{2.15}
\ene
Note that $ \Delta$ diverges at the boundary of $K$.

We take here the  {\it material interpretation} and we consider our
transformation as a bijection between two different spaces,
$\Omega_0$ and $\Omega$. However, our transformation can be
considered, as well, as a change of coordinates in $\Omega_0$. These two points of view are mathematically equivalent. This
means that under our transformation  Maxwell
equations in $\Omega_0$ and in $\Omega$ have the same
invariance  that they have under change of coordinates in
three-space. See, for example, \cite{po}. Let us denote by
$\e_0,\h_0,\b_0, \d_0,  \varepsilon^{\lambda\nu}_0,
\mu^{\lambda\nu}_0$, respectively, the electric and magnetic fields,
the magnetic induction, the electric displacement, and the
permittivity and permeability of $\Omega_0$.  $
\varepsilon^{\lambda\nu}_0, \mu^{\lambda\nu}_0$, are positive
Hermitian matrices that are constant in $\Omega_0$.
The electric field is a covariant vector that transforms as,

\beq E_\lambda(\x) = A_\lambda^{\lambda'}(\y)E_{0,\lambda'}(\y).
\label{2.16} \ene

The magnetic field $\h$ is a covariant pseudo-vector, but as we only
consider space transformations with positive determinant, it also
transforms as in (\ref{2.16}). The magnetic induction $\b$ and the
electric displacement $\d$ are contravariant vector densities of
weight one that transform as

\beq B^\lambda(\x) =  \left(\Delta (\y)\right)^{-1}
A_{\lambda'}^{\lambda}(\y) B^{\lambda'}_0(\y),
\label{2.17}
\ene
with the same transformation for $\d$. The permittivity and
permeability are contravariant tensor densities of weight one that
transform as,
\beq
\varepsilon^{\lambda\nu}(\x)=  \left(\Delta
(\y)\right)^{-1} A^{\lambda}_{\lambda'}(\y)\, A^{\nu}_{\nu'}(\y)\,
\varepsilon^{\lambda' \nu'}_0(\y),
\label{2.18}
\ene
with the same transformation for $ \mu^{\lambda\nu}$.  Maxwell equations
(\ref{2.1}-\ref{2.4}) are the same in both spaces $\Omega$ and
$\Omega_0$. Let us denote by $\varepsilon_{\lambda \nu},
\mu_{\lambda \nu}, \varepsilon_{0\lambda \nu}, \mu_{0\lambda \nu}$,
respectively, the inverses of the corresponding permittivity and
permeability. They are covariant tensor densities of weight minus
one that transform as,

\beq \varepsilon_{\lambda\nu}(\x)=  \Delta (\y)
A^{\lambda'}_{\lambda}(\y)\, A^{\nu'}_{\nu}(\y)\,
\varepsilon_{0\lambda' \nu'}(\y),\, \mu_{\lambda\nu}(\x)=  \Delta
(\y) A^{\lambda'}_{\lambda}(\y)\, A^{\nu'}_{\nu}(\y)\,
\mu_{0\lambda' \nu'}(\y).
\label{2.19}
\ene
 We have that
\beq
 \det \varepsilon^{\lambda\nu}=\Delta^{-1} \det \varepsilon^{\lambda\nu}_0,\,
 \det \mu^{\lambda\nu}=\Delta^{-1} \det \mu^{\lambda\nu}_0,\,
\label{2.20}
\ene

\beq
\det \varepsilon_{\lambda\nu}=\Delta \det
\varepsilon_{0\lambda\nu},\,  \det \mu_{\lambda\nu}=\Delta \det
\mu_{0\lambda\nu}.
\label{2.21}
\ene

 The matrices $ \var^{\lambda \nu}, \mu^{\lambda \nu}$ are degenerate at $\partial K$ and  the matrices
 $ \var_{\lambda \nu}, \mu_{\lambda \nu}$ are singular at $\partial K$.

As the $\varepsilon^{\lambda\nu}$ and $\mu^{\lambda\nu}$ are degenerate at
the boundary of the cloaked object $K$, we have to make precise what do we mean by a solution to
Maxwell equations in neighborhood of $\partial K$. In other words, we have to specify the {\it cloaking
boundary conditions} that the solutions have to satisfy on the outside and on the inside of $\partial K$.
We solved this problem in \cite{we1,we2} by requiring that the fundamental principle of energy conservation
be satisfied. That is to say, we obtained the appropriate boundary conditions by requiring that the solutions
conserve energy. Note that as our media are loss-less  energy has to be conserved. Furthermore, any energy loss
of the incoming waves could {\it a priori} be detected, and, in consequence, energy conservation is essential
for cloaking purposes. We briefly review the results of \cite{we1,we2}.

We first consider the problem in
$\Omega$. We write Maxwell equations in Schr\"odinger form. For this purpose we denote  by
$ {\mathbf \var}$ and $ {\mathbf \mu}$, respectively, the matrices with entries
$\var_{\lambda\nu}$ and $\mu_{\lambda\nu}$.  Recall that $\left(\nabla\times \e\right)^\lambda=
s^{\lambda\nu\rho} \frac{\partial}{\partial x_\nu }E_\rho$, where $s^{\lambda\nu\rho}$ is the permutation
contravariant pseudo-density of weight $-1$
(see section 6 of chapter II of \cite{po}, where a different notation is used).

We  define the following formal differential operator,
\beq
a_\Omega \left(\begin{array}{c}\e\\ \h\end{array}\right)=i \left(\begin{array}{c}
{\mathbf\var} \nabla\times \h\\- {\mathbf \mu} \nabla\times \e\end{array}\right).
\label{2.22}
\ene
Here, as usual, we denote, $\mathbf\var \nabla\times \h:= \var_{\lambda\nu} (\nabla \times \h)^\nu$, and
$\mathbf\mu \nabla\times \e=\mu_{\lambda\nu}
(\nabla \times \e)^\nu$.

Equation (\ref{2.1}) is equivalent to,
\beq
i\frac{\partial}{\partial t}\left(\begin{array}{c}\e\\\h\end{array}\right)= a_\Omega\left(\begin{array}{c}\e\\\h\end{array}\right),
\label{2.23}
\ene

and equation (\ref{2.3}) is equivalent to
\beq
-\omega\left(\begin{array}{c}\e\\\h\end{array}\right)= a_\Omega\left(\begin{array}{c}\e\\\h\end{array}\right).
\label{2.24}
\ene
Note that since the matrices $\mathbf \var, \mu$ are singular at $\partial \Omega$ the operator $a_\Omega$ has
coefficients that are singular  at $\partial \Omega$. This is the reason why we have to be careful when
defining the solutions.

It is necessary to define  equation (\ref{2.22}) in an appropriate linear subspace of the Hilbert space of all
finite energy fields in $\Omega$. We designate  by $\H_{\Omega E}$ the Hilbert space of
all measurable, $\CE^3-$ valued functions defined on $\Omega$  that are square integrable with the weight
$\var^{\lambda\nu}$ and the scalar product,
\beq
\left(\e^{(1)}, \e^{(2)}\right)_{\Omega E}:= \int_{\Omega}E^{(1)}_\lambda\,\var^{\lambda\nu}\,
\overline{E^{(2)}_\nu}\, d\x^3.
\label{2.25}
\ene
Moreover,  we denote  by $\H_{\Omega H}$ the Hilbert space of all
measurable, $\CE^3-$ valued functions defined on $\Omega$  that are square integrable with the weight
 $\mu^{\lambda\nu}$ and the scalar product,
\beq
\left(\h^{(1)}, \h^{(2)}\right)_{\Omega H}:= \int_{\Omega}H^{(1)}_\lambda\,\mu^{\lambda\nu}\,
\overline{H^{(2)}_\nu}\, d\x^3.
\label{2.26}
\ene

The Hilbert space of finite energy fields in $\Omega$ is the direct sum

\beq
\H_\Omega:= \H_{\Omega E}\oplus \H_{\Omega H}.
\label{2.27}
\ene
We first define $ a_\Omega$ in a nice set of functions where it makes sense, that we take as $\C^1_0(\Omega)$.
In physical terms this means that we start with the minimal assumption that Maxwell's equations are satisfied
in classical sense away from the boundary of $\Omega$.
 $a_\Omega$ with domain $D(a_\Omega):= \C^1 _0(\Omega)$ is a symmetric operator in $\H_\Omega$, i.e.
$ a_\Omega \subset a_\Omega^\ast$.
To construct a unitary dynamics that preserves energy we have to analyze the self-adjoint extensions of
$a_\Omega$, what in physical terms means that we have to make precise in what  sense Maxwell's equations are
solved up to $\partial \Omega$. In other words,  to construct finite-energy solutions of (\ref{2.23}),
with constant energy we have to demand that the initial finite energy fields , $ (\e(0), \h(0))^T$  belong to
the domain of one of the self-adjoint extensions of $a_\Omega$.  The key issue is that $a_\Omega$ has only one
self-adjoint extension (i.e. it is essentially self-adjoint) that we denote by $A_\Omega$. Moreover, $A_\Omega$
is unitarily equivalent to the free Maxwell propagator, $A_0$, in $\Omega_0$. The unitary equivalence is
generated  by (\ref{2.16}) and by the same transformation for the magnetic field. This means that there
is only one dynamics in $\Omega$ that preserves energy, and that this dynamics is generated by $A_\Omega$.
As $A_\Omega$ and $A_0$ are unitarily equivalent, the dynamics that they   generate are  physically equivalent,
and this is the deep reason, from the point of view of fundamental physics, why there is perfect cloaking of
passive and active devices.

Solutions to (\ref{2.3}, \ref{2.4}) in general do not have finite energy  because they do not have enough
decay at infinity to be square integrable over all $\Omega$. Then, we only require  that they are of locally
 finite energy in the sense that the  electric  and the magnetic fields are square integrable over
 every bounded subset of $\Omega$, respectively, with the weight
 $\var^{\lambda\nu}$, and $\mu^{\lambda\mu}$.
 Moreover, in order that the problem (\ref{2.3}, \ref{2.4}) is well-posed -in the sense that it is
 self-adjoint-
 the solutions with locally finite energy have to be locally in the domain of the only self-adjoint extension
 of $a_\Omega$, that is to say,  they  have to be in the domain of $A_\Omega$ when multiplied by
 any continuously differentiable function with
support in a bounded subset of $\overline{\Omega}$.

On the basis of these considerations we proved in \cite{we1,we2}
that the solutions with locally finite energy in $\Omega$ are solutions in distribution sense to
(\ref{2.3}, \ref{2.4}) that satisfy,
\beq
\int_{O}E_\lambda\,\var^{\lambda\nu}\, \overline{E_\nu}\, d\x^3 +
\int_{O}H_\lambda\,\mu^{\lambda\nu}\, \overline{H_\nu}\, d\x^3 <
\infty,
\label{2.28}
\ene
 for every bounded set $O \subset \Omega$. Moreover, they have to satisfy the
 {\it cloaking boundary condition},

\beq
\e\times \mathbf n=0, \h\times  \mathbf n=0, \,\, \hbox {\rm in }\,\, \partial \Omega=\partial K_+,
\label{2.29}
\ene

where $\partial K_+$ is the outside of the boundary of the cloaked object and $\mathbf n$ is the normal
vector to $\partial K_+$.

Note that as $A_\Omega$ is the only self-adjoint extension of $a_\Omega$,  this is the only possible
self-adjoint boundary condition on $\partial K_+$. It is self-adjoint because the matrices
$\var, \mu$  are singular at $\partial K_+$. Hence, cloaking as boundary value problem consists of
 finding
a solution to (\ref{2.3}, \ref{2.4}) in $\Omega$  with locally-finite energy
that satisfies the {\it cloaking boundary condition}  given in (\ref{2.29}).

\bull

Let us now consider the propagation of electromagnetic  waves inside the cloaked object.
We assume that in each $K_j$ the permittivity and the permeability are given
by $\var^{\lambda\nu}_j, \mu^{\lambda\nu}_j$, with
inverses $\var_{j\lambda\nu}, \mu_{j\lambda\nu}$ and where $\var_j, \mu_j$ are the matrices with entries
$\var_{j\lambda\nu}, \mu_{j\lambda\nu}$.
Furthermore, we assume that $  0 < \var^{\lambda\nu}_j(\x), \mu^{\lambda \nu}_j(\x) \leq C, \x \in K_j$
and that
for any compact set $Q$ contained in the interior
of $K_j$ there is a positive constant $C_Q$ such that $\det \var^{\lambda\nu}_j(\x) > C_Q,
\det\mu^{\lambda \nu}_j(\x) > C_Q, \x \in Q,j =1,2,\cdots,N$. In other words, we only
allow for possible singularities of $\var_j, \mu_j$ on the boundary of $K_j$.

We designate  by $\H_{j E}$ the Hilbert space of all
measurable, $\CE^3-$ valued functions defined on $K_j$  that are square integrable with the weight
 $\var^{\lambda\nu}_j$ and the scalar product,
\beq
\left(\e^{(1)}_j, \e^{(2)}_j\right)_{j E}:= \int_{K_j}E^{(1)}_{j\lambda}\,\var^{\lambda\nu}_j\,
\overline{E^{(2)}_{j\nu}}\, d\x^3.
\label{2.30}
\ene

 Similarly, we denote  by $\H_{j H}$ the Hilbert space of all
measurable, $\CE^3-$ valued functions defined on $K_j$  that are square integrable with the weight
$\mu^{\lambda\nu}_j$ and the scalar product,
\beq
\left(\h^{(1)}_j, \h^{(2)}_j\right)_{j H}:= \int_{K_j}H^{(1)}_{j\lambda}\,\mu^{\lambda\nu}_j\,
\overline{H^{(2)}_{j\nu}}\, d\x^3.
\label{2.31}
\ene

The Hilbert space of finite energy fields in $K_j$ is the direct sum

\beq
\H_{j}:= \H_{j E}\oplus \H_{j H},
\label{2.32}
\ene
and the Hilbert space of finite energy fields in the cloaked object $K$ is the direct sum,

$$
\H_K:=  \oplus_{j=1}^N \H_{j}.
$$

The complete Hilbert space of finite energy fields  including the cloaked object is,

\beq
\H:= \H_\Omega\oplus \H_K.
\label{2.33}
\ene

We now write (\ref{2.1}) as a Schr\"odinger equation in each $K_j$ as before.
We define the following formal differential operator,
\beq
a_j \left(\begin{array}{c}\e_j\\ \h_j\end{array}\right)=i \left(\begin{array}{c}
\var_j \nabla\times \h_j\\- \mu_j \nabla\times \e_j\end{array}\right).
\label{2.34}
\ene

Equation (\ref{2.1}) in $K_j$ is equivalent to
\beq
i\frac{\partial}{\partial t}\left(\begin{array}{c}\e_j\\\h_j\end{array}\right)= a_j\left(
\begin{array}{c}\e_j\\\h_j\end{array}\right).
\label{2.35}
\ene
Let us denote the interior of
$K_j$  by $\stackrel{o}K_j:= K_j \setminus \partial K_j$.
 Then, $a_j$ with domain $C^1_0(\stackrel{o}K_j)$ is a symmetric operator in $\H_{j}$.
We denote,

\beq
a:= a_\Omega\oplus a_K, \, \hbox{\rm where}\, a_K:= \oplus_{j=1}^N a_j,
\label{2.36}
\ene
with domain,
\beq
D(a):= \left\{ \left(\begin{array}{c}\e_\Omega\\ \h_\Omega\end{array}\right)\oplus_{j=1}^N
\left(\begin{array}{c}\e_j\\ \h_j\end{array}\right)
\in \C^1_0(\Omega)\oplus_{j=1}^N \C^1_0(\stackrel{o}K_j)\right\}.
\label{2.37}
\ene
The operator $a$ is symmetric in $\H$. The possible unitary dynamics that preserve energy for the whole
system, including the cloaked object, $K$, are given by the self-adjoint extensions of $a$.
We  proved in \cite{we1,we2} that
every self-adjoint extension, $A$, of $a$ is the direct sum of  $A_\Omega$ and of some self-adjoint
extension, $A_K$, of $a_K$, i.e.,
\beq
A= A_\Omega \oplus A_K.
\label{2.38}
\ene
This result implies  that the cloaked object $K$ and the exterior $\Omega$ are completely decoupled. That
 electromagnetic waves outside  $K$ can not go inside and vice versa that waves inside  can not propagate
outside, and that there is perfect cloaking of passive and active devices. Choosing a particular  self-adjoint
 extension $A_K$ amounts to fixing a boundary condition in the inside of the
boundary of the cloked object, $\partial K_-$. The self-adjoint
extension -or boundary condition- that nature will take  depends  on
the properties of the media inside the cloaked object. Note that
this does not mean that we have to put any physical surface, a
lining, on the surface of the cloaked object to enforce any
particular boundary condition  on the inside, since this plays no
role in  the cloaking outside. It is, however, of importance to
determine what the self-adjoint extension in $K$, or the interior
boundary condition, has to be for specific cloaked objects. See
\cite{we1,we2} for a detailed discussion of these issues.

The problem that we address on this paper is to determine what the boundary conditions in $\partial K_-$ have
to be in the particular case where the permittivity and the permeability  in $K$  are bounded and non-degenerate,
i.e., when the matrices $\var^{\lambda\nu}_j,\mu^{\lambda \nu}_j$ are bounded above and below in $K_j$,

\beq
0 < C_1 <  \var^{\lambda\nu}_j, \mu^{\lambda \nu}_j < C_2 , \x \in K_j, j=1,2,\cdots,N,
\label{2.39}
\ene
for some positive constants $C_1,C_2$. This is clearly the most important case in the applications. It
corresponds to a  standard  object that is cloaked with a metamaterial.

Let us consider the case of an active
device with electric and magnetic currents in $K$. The Maxwell equations at frequency $\omega$ are,

\beq
 \nabla \times \h \,=\,i\omega \d+ J,
 \label{2.40}
\ene
\beq
\nabla \times \e =  -i\omega \b- J_m,
\label{2.41}
\ene

where $J$ and $J_m$ are, respectively, the electric and the magnetic currents, that we assume are different
from zero only in $K$.

As we mentioned above, we have already proven in \cite{we1,we2} that
energy conservation implies that the electromagnetic waves inside
$K$ can not propagate outside and that, vice versa, the waves
outside  can not go inside. The key issue here is that this is
consistent with Maxwell equation (\ref{2.40}) only if the normal
component of the total current (i.e. the sum of the displacement
current and the electric current)  vanishes at $\partial K_-$,
i.e., if

\beq
\left(i\omega \d+ J\right)\cdot \mathbf n|_{\ds \partial K_{-}}=0,
\label{2.42}
\ene
where as usual by $\partial K_{-}$ we mean that we approach the boundary  from the inside.
 In a similar way, the consistency with Maxwell equation (\ref{2.41}) implies that,

\beq
\left(i\omega \b+ J_m\right)\cdot \mathbf n|_{\ds \partial K_{-}}=0.
\label{2.43}
\ene

Note that we do not need to ask that (\ref{2.42}, \ref{2.43}) hold at $\partial K_+$ because as we assume that $J,J_m$ are identically zero
outside $K$  conditions (\ref{2.42}, \ref{2.43}) in
$\partial K_+$ follow from  equations (\ref{2.14}, \ref{2.15},
\ref{2.17}) and the same transformation equation for $\d$, since the solution in $\Omega$ is obtained applying the
transformation formulae to a solution in $ \ere^3_0$ \cite{we1,we2}. Moreover, the boundary conditions
(\ref{2.42}, \ref{2.43}) on $\partial K_-$ and Maxwell equations (\ref{2.40}, \ref{2.41}) imply that,

\beq
\left( \nabla \times \h\right)\cdot \mathbf n|_{\ds \partial K_{-}}=0, \, \left( \nabla \times \e\right)
\cdot \mathbf n|_{\ds \partial K_{-}}=0.
\label{2.45}
\ene
Hence, we have proven that the boundary conditions that we have to impose on the inside of the boundary of the
cloaked object are (\ref{2.45}), namely that the normal components of the curl of the electric and the magnetic
fields have to vanish. The Maxwell propagator $a_K$ with the boundary condition (\ref{2.45}) has already been
studied in the mathematical literature, in particular in relation with Beltrami fields. As it was to be expected
from our analysis,  $a_K$ with the boundary condition (\ref{2.45}) is a self-adjoint operator. In other words,
the boundary condition (\ref{2.45}) defines the self-adjoint realization, $A_K$, of the Maxwell propagator in $K$
that is imposed by energy conservation. For the proof of self-adjointness, as  well as other issues, including
the formulation of (\ref{2.45}) in weak sense see \cite{pi1,pi2,pi3,fi}, in particular, see page 158, Theorem 2.1,
Corollary 2.1.1, page 164, and Theorem 2.3 of \cite{pi3}. Note that in our case the Neumann fields are zero.
 Remark that imposing that the normal components of $D$ and $B$ are zero at $\partial K_-$ is not a self-adjoint boundary condition, i.e.,
 it does not define a self-adjoint extension of Maxwell generator in $K$.

We have now a complete formulation of cloaking as a boundary value problem. It consists of finding a solution
of Maxwell equations (\ref{2.40}, \ref{2.41}) in distribution sense in $\ere^3 \setminus \partial K$,
with locally finite energy, i.e., they satisfy

\beq
\int_{O}E_\lambda\,\var^{\lambda\nu}\, \overline{E_\nu}\, d\x^3 +
\int_{O}H_\lambda\,\mu^{\lambda\nu}\, \overline{H_\nu}\, d\x^3 <
\infty,
\label{2.46}
\ene
where $O$ is any bounded subset of $\ere^3$. Moreover, they have to satisfy the  {\it cloaking
boundary conditions}

\beq
\e\times \mathbf n=0, \h\times  \mathbf n=0, \,\, \hbox {\rm at }\,\, \partial \Omega=\partial K_+,
\label{2.47}
\ene
and

\beq
 \left(\nabla \times \e\right)\cdot \mathbf n
=0,\, \left(\nabla \times \h\right)\cdot \mathbf n=0, \, \, \, \hbox{at}\, \partial K_-.
\label{2.48}
\ene

We have derived the boundary conditions (\ref{2.47},  \ref{2.48}) by requiring that the solutions to the fixed frequency Maxwell equations
(\ref{2.40}, \ref{2.41}) are (locally) in the domain of the appropriate self-adjoint extension  (\ref{2.38}) of Maxwell generator (\ref{2.36}).
Note that when we define the self-adjoint operator $A_K$ we have to require that all functions on its domain satisfy the boundary conditions
(\ref{2.48}), not just the solutions to the fixed frequency Maxwell equations. In fact, the choice of the self-adjoint  Maxwell generator $A_K$
has implications that go well beyond the formulation of cloaking as a boundary value problem. For example, it determines the time evolution of
finite-energy wave packets in the time domain. See \cite{we1,we2} for this issue.

\section{The Case of a Radial Source at the  Boundary of $\mathbf K$}\sss

In this section we illustrate our method by considering  an active device given by a
radial electric current  at the  boundary of $K$. The case of a magnetic current at the boundary
of $K$
follows in the same way. We assume that $K$ is isotropic and spherically stratified, i.e., that the permittivity and the permeability
depend only on $|\x|$.
The case where $K$ is isotropic and homogeneous, and with  an electric dipole contained in the interior of K
was already considered in \cite{zcwk}. We verify
in this particular case,  by an explicit computation, that our {\it
cloaking boundary conditions} are satisfied and that cloaking of
active devices holds, even if the current is at the boundary of the
cloaked object, as we have proven in Section 2 in the general case
where there is no explicit solution.

We assume that we have only one spherical cloak, $K$, located at the origin, i.e., $N=1, \mathbf c_1=0$,
that $K$ is
isotropic with   permittivity and permeability, $\var_1, \mu_1$, that are bounded, that they have  a positive
lower bound, and that
they depend only on $r:=|\x|$. For simplicity we take a first order transformation with
$ g(\rho)= \frac{b-a}{b}\rho +a,0 < a < b$. In the cloaking layer the permittivity and the
permeability tensors are given by \cite{pss1,sps},

\beq
 \var^{\lambda\nu}= \var_r \hat{\x}^\lambda
\hat{\x}^\nu+ \var_t
\hat{\mathbf \theta}^\lambda \hat{\mathbf \theta}^\nu + \var_t \hat{\mathbf\varphi}^\lambda
\hat{\mathbf\varphi}^\nu,\mu^{\lambda\nu}=
\mu_r \hat{\x}^\lambda \hat{\x}^\nu+ \mu_t
\hat{\mathbf \theta}^\lambda \hat{\mathbf \theta}^\nu + \mu_t \hat{\mathbf\varphi}^\lambda
\hat{\mathbf\varphi}^\nu, a < |\x| < b,
\label{3.1}
\ene
where,
$ \hat{\mathbf\theta}, \hat{\mathbf\varphi}$ are unit tangent vectors, respectively, to the coordinate lines,
$ r, \varphi$
constant and $ r, \theta$ constant, in spherical coordinates, $r, \theta, \varphi$.
Moreover,
\beq
\var_t/\var_0= \mu_t/\mu_0 = b/(b-a), \var_r/\var_t= \mu_r/\mu_t= (r-a)^2/r^2,  a < r < b.
\label{3.2}
\ene
 We assume that for
$ r > b$ the medium is homogeneous and isotropic with permittivity and permeability, $ \var_0,\mu_0$.

The expression of the transverse electric, TE, and transverse magnetic, TM,
fields in terms of potentials given in Section 8.6 of \cite{vb} remain true in our case
(remark the $\var_t$ and $\mu_t$ are constant in the cloaking layer). See also
\cite{cwzk}, \cite{zcwk}. TE and TM fields decouple, and since we have a radial electric
current we only consider TM modes. Assuming that $ J=J_r(r) \hat{\x}$ and that $J_m=0$, the TM fields are
given by the potential as follows \cite{vb},
\begin{eqnarray}
\e_t&=& \frac{-i}{\omega \var_t r}\grad \frac{d}{dr}I, \nonumber \\
\h_t&=& \frac{1}{ r}\left( \grad I \times \hat{\x}\right),\nonumber \\
\e_r&=&  - \frac{1}{i\omega \var_r r^2} \Delta_{\theta\varphi} I-\frac{1}{i\omega  \var_r} J_r.
\label{3.3}
\end{eqnarray}
We expand the potential in spherical harmonics,
$$
I= \sum_{mn} I(r)_{mn}  Y^m_n(\theta, \varphi),
$$
and we assume that the radial current has the following expansion,
$$
J_r= \sum_{mn} J_{m n} \,\delta(r-a)\, Y^m_n(\theta, \varphi),
$$
for some constants $J_{mn}$.  For example, for an electric dipole located at $(0,0,a)$  we have that
$J_{m n}= - \frac{i \omega P_e}{2\pi a^2} \sqrt{\frac{2n+1}{4\pi}}\delta_{m,0}$ \cite{vb}.

We now set the inner boundary at $a+\delta$, for small $ \delta >0$, i.e., we assume that the permittivity and permeability are equal to $ \var_1, \mu_1$ for $r < a+\delta$,
that they are given by (\ref{3.1}, \ref{3.2}) for $ a+\delta < r < b$ and by $\var_0, \mu_0$ for  $ r > b$. We  compute the solution, and then
 we take the limit
as $ \delta$ tends to zero (\cite{rynq}).

For $ 0 < r < a+\delta$, the potential $I_{a,mn}$ created by the source satisfies the following equation \cite{vb}, where we denote
$\frac{d}{dr}$ by $'$,
\beq
I_{a,mn}''- \frac{1}{\var_1} \var_1' \, I_{a,mn}'+\left[\omega^2 \var_1\mu_1 - \frac{n(n+1)}{r^2}\right]
I_{a,mn}=- J_{mn} \delta(r-a).
\label{3.4}
\ene
Let $v_{n}$ and $w_{n}$ be to independent solutions of the homogeneous equation with $v_{n}$ regular at
zero. For example, if $\var_1(r),\mu_1(r)$ are piecewise constant, $v_{n}$ can be taken as a
Ricatti-Bessel function of the first kind
 and $w_{n}$ as a Ricatti-Bessel function of the second or third kind \cite{vhu}
 in each layer where $\var_1,\mu_1$ are
constant. Hence, the solution to (\ref{3.4}) is given by,

\beq
I_{a, mn}(r)=  \frac{J_{mn}}{v_{n}'(a)
w_{n}(a)-v_{n}(a) w_{n}'(a)} \left\{\begin{array}{l}w_{n}(a) v_{n}(r), 0 < r < a, \\ \\
 v_{n}(a) w_{n}(r), a < r <
 a+\delta.
 \end{array} \right.
 \label{3.5}
 \ene

We suppose that $ v_{n}(a)$ and $w_{n}(a)$ are different from zero.
We assume that there is also a reflected wave. Then, the total potential is given by,
\beq
I_{mn}=I_{a,mn}(r)+ R_{mn} v_{n}(r), 0 < r < a+\delta,
\label{3.6}
\ene
where the $R_{mn}$ are  the reflection coefficients.
In the cloaking layer the potential satisfies the equation (remark that $\var_t,\mu_t$ are constant),

\beq
I_{mn}''+\left[\omega^2 \var_t\mu_t - \frac{\var_t}{\var_r}\frac{n(n+1)}{r^2}\right]
I_{mn}=0, a+\delta < r <b.
\label{3.7}
\ene
The solution is given by Ricatti-Bessel functions of the first and second kind,

\beq
I_{m n}(r)= c_{mn} \psi_n( k_t(r-a) )+d_{mn} \chi_n(k_t(r-a)), k_t:= \omega \sqrt{\var_t\mu_t},\,
a+\delta < r < b.
\label{3.8}
\ene
Outside of the cloaking layer the potential satisfies equation (\ref{3.7}) with $\var_t=\var_r=\var_0$ and
$\mu_t=\mu_0$. The solution is an outgoing wave,
\beq
I_{mn}(r)= T_{mn} \,\zeta_n(k_0 r), k_0:= \omega \sqrt{\var_0 \mu_0}, b < r,
\label{3.9}
\ene
where $\zeta$ is a Ricatti-Bessel function of the third kind.

Requiring that the tangential components of the electric and the magnetic fields are continuous at $ a+\delta$
and $b$ we obtain the following equations,

\begin{eqnarray}
\frac{1}{\var_1(a+\delta)}\left( I_{a,mn}'(a+\delta)+ R_{mn} v_{n}'(a+\delta)\right)&=&
\frac{k_t}{\var_t}\left( c_{mn} \psi_n'(k_t\delta)+ d_{mn} \chi_n'(k_t\delta)\right),\label{3.10}\\
 I_{a,mn}(a+\delta)+ R_{mn} v_{n}(a+\delta)&=&
 c_{mn} \psi_n(k_t\delta)+ d_{mn} \chi_n(k_t\delta),\label{3.11} \\
\frac{k_t}{\var_t}\left( c_{mn} \psi_n'(k_t (b-a))+ d_{mn} \chi_n'(k_t(b-a)\right)&=& \frac{k_0}{\var_0}
 T_{mn}  \zeta_n'(k_0 b), \label{3.12}
\\
 c_{mn} \psi_n(k_t (b-a))+ d_{mn} \chi_n(k_t(b-a))&=&  T_{mn}  \zeta_n(k_0 b).
\label{3.13}
\end{eqnarray}

Let us denote,
\beq
\alpha:= \left(\frac{\mu_t\var_0}{\mu_0\var_t}\right)^{1/2},
\label{3.14}
\ene
\beq
\beta_1:= \zeta_n'(k_0 b) \chi_n(k_t(b-a)) -\alpha \zeta_n(k_0 b) \chi_n'(k_t(b-a)),
\label{3.15}
\ene
\beq
\beta_2:=  \zeta_n'(k_0 b) \psi_n(k_t(b-a)) -\alpha \zeta_n(k_0 b) \psi_n'(k_t(b-a)), \, \beta_3:=
 - \frac{\beta_1}{\beta_2},
\label{3.16}
\ene
\beq
\beta_4:= v_{n}'(a+\delta) (\beta_3 \psi_n(k_t \delta)+\chi_n(k_t\delta))- \frac{k_t \var_1(a+\delta)}{\var_t}v_{n}(a+\delta)
(\beta_3 \psi_n'(k_t \delta)+\chi_n'(k_t\delta)),
\label{3.17}
\ene

Solving (\ref{3.10}-\ref{3.13}) in the generic case where $\beta_2\neq 0$ we obtain that,

\beq
R_{mn}= \frac{1}{v_n(a+\delta)}\left(( \beta_3 \psi_n(k_t\delta)+ \chi_n(k_t\delta)) d_{mn}-I_{a,mn}(a+\delta) \right) ,
\label{3.18}
\ene

\beq
c_{mn}= \beta_3 d_{mn},
\label{3.19}
\ene
\beq
d_{mn}= J_{mn} v_{n}(a)\frac{1}{\beta_4},
\label{3.20}
\ene
\beq
T_{mn}= \frac{1}{\zeta_n(k_0b)}  (\beta_3 \psi_n(k_t (b-a))+ \chi_n(k_t(b-a)) d_{mn}.
\label{3.21}
\ene

Using the expansions of the spherical Bessel functions for small argument given in equations (10.1.2, 10.1.3)
of \cite{as} we prove that,

\beq
d_{mn}= - J_{mn} \frac{v_n(a)\var_t }{ k_t\var_1(a+\delta)v_n(a+\delta) n (2n-1)!!} (k_t\delta)^{n+1}+O((k_t \delta)^{n+2}).
\label{3.22}
\ene
Using (\ref{3.22}) we obtain the small $\delta$ expansions of $R_{mn}, c_{mn}, T_{mn}$, respectively, from
(\ref{3.18}, \ref{3.19}, \ref{3.21}).

\beq
R_{mn}= \frac{1}{v_n(a+\delta)}( -I_{a,mn}(a+\delta)+O(k_t\delta),
\label{3.23}
\ene
\beq
c_{mn}= \beta_3 \left(- J_{mn} \frac{v_n(a)\var_t }{k_t\var_1(a+\delta)v_n(a+\delta) n (2n-1)!!} (k_t\delta)^{n+1}
+O((k_t \delta)^{n+2})\right),
\label{3.24}
\ene
\begin{eqnarray}
T_{mn}&=&  -\frac{J_{mn} v_n(a)\var_t(\beta_3 \psi_n(k_t (b-a))+ \chi_n(k_t(b-a))}{k_t \var_1(a+\delta)\zeta_n(k_0b)
v_n(a+\delta) n (2n-1)!! }
         (k_t\delta)^{n+1}
        +O((k_t \delta)^{n+2}).
\label{3.25}
\end{eqnarray}
Moreover, by (\ref{3.6}, \ref{3.11}, \ref{3.22}, \ref{3.24}),
\beq
I_{mn}(a+\delta)=J_{mn} \frac{v_n(a)\var_t}{v_n(a+\delta) n k_t \var_1(a+\delta)} k_t\delta+ O((k_t\delta)^2).
\label{3.26}
\ene
It follows that  when $\delta=0, c_{mn}=d_{mn}=T_{mn}=0$, and the electric and magnetic fields
outside $K$ are zero, as predicted by our theoretical results. Also, for $\delta=0$, $I_{mn}(a)=0$, and then,
by equation (\ref{3.3}) the radial component of the sum of the displacement current and the electric current
vanishes at $r=a$. By Maxwell equation (\ref{2.40}) the normal component  of the curl of the magnetic field
vanishes ar $r=a$. The corresponding statement for the normal component of the curl of the electric field is
trivial in this case  by Maxwell equation (\ref{2.41}) and as the magnetic field is transversal and
the magnetic current is zero. Hence, the boundary conditions (\ref{2.45}) are satisfied and cloaking holds even if the current is at the
boundary of $K$, as predicted by our
theoretical results.

\section{Conclusions}\sss
The results of this paper and of \cite{we1,we2}
give a complete rigorous mathematical analysis of point transformed electromagnetic invisibility cloaks.
We solved the mathematical challenges posed by the fact that the permittivity and the permeability are degenerate at the boundary of the cloaked
object $K$.
In particular,  we characterized all possible ways to define solutions of Maxwell equations  that are compatible with energy conservation. This result
was obtained by characterizing all possible boundary conditions at $\partial K_\pm$ that are allowed by
energy conservation.  They correspond to all self-adjoint extensions of the Maxwell generator. As it turned out, all self-adjoint extensions are the
direct sum of some self-adjoint extension inside $K$ with a fixed self-adjoint extension outside $K$. This implies that the solutions inside and
outside of $K$ are completely decoupled from each other and that, in general, they are  discontinuous at $\partial K$.
We also proved   that cloaking of passive and active devices always holds for all possible ways to define solutions that
satisfy energy conservation, i.e., with self-adjoint boundary
conditions. The boundary condition at $\partial K_+$ is always that the tangential components  of both the electric
and the magnetic field vanish. At $\partial K_-$ the boundary condition can by any self-adjoint boundary condition for the Maxwell generator in $K$.
The particular self-adjoint boundary condition that nature will take depends on the specific properties of the media inside $K$.
In this paper we solved the problem of determining  the appropriate boundary conditions at $\partial K_-$ in the
 case where the permittivity and the permeability inside $K$ are bounded  and
have a positive lower bound. This corresponds to the situation where
we have a standard object $K$ -that could be anisotropic and
inhomogeneous, but whose permittivity and permeability  are neither
singular nor degenerate- that is coated by a transformation medium
that is degenerate at $\partial K_+$. We also allow for passive and active
devices in $K$. This is perhaps the more important case in the
applications.

In this way we have obtained a complete formulation of cloaking with passive and
active devices as a boundary value problem.
It consists of finding a solution
of Maxwell equations (\ref{2.40}, \ref{2.41}) at frequency $\omega$, in distribution sense in $\ere^3 \setminus \partial K$,
with locally finite energy, i.e., they satisfy

$$
\int_{O}E_\lambda\,\var^{\lambda\nu}\, \overline{E_\nu}\, d\x^3 +
\int_{O}H_\lambda\,\mu^{\lambda\nu}\, \overline{H_\nu}\, d\x^3 <
\infty,
$$
where $O$ is any bounded subset of $\ere^3$. Moreover, they have to satisfy the following {\it cloaking
boundary conditions}

\beq
\e\times \mathbf n=0, \h\times  \mathbf n=0, \,\, \hbox {\rm at }\,\, \partial K_+,
\label{4.1}
\ene
and

\beq
  \left(\nabla \times \e\right)\cdot \mathbf n
=0,\, \left(\nabla \times \h\right)\cdot \mathbf n=0, \, \, \, \hbox{at}\, \partial K_-.
\label{4.2}
\ene

In the case of one spherical cloaked object, that is isotropic and spherically stratified, and that  has an active device given by a
radial electric current  at the boundary,  we verified by an  explicit computation that our {\it
cloaking boundary conditions} are satisfied and that cloaking of active devices holds, even if the current is at the boundary of the
cloaked object.

A novel aspect of our work is that  we proved our results for  transformations media that are obtained from
general anisotropic media, i.e.,   that it is not necessary to transform from isotropic media.
This means that it is possible to cloak  objects  that are contained inside general anisotropic materials, general crystals for example.
For this purpose, we just have to take as the permittivity and the permeability of the general anisotropic medium before transformation the ones of the
general anisotropic material, or the general crystal, that contains the object that we wish to cloak.
The fact that it is possible to cloak objects inside general anisotropic media opens the way to other interesting potential
applications, for example to guide electromagnetic waves under quite general circumstances.

The results above -that are proven for cloaks in  exact transformation media (ideal cloaks)- set the stage for the rigorous study of the cloaks in
the approximate transformation media that one has to consider in practical situations, what is one of the main open questions on this area. This issue
can be understood as the problem of the stability of cloaking under the perturbation on the permittivity and the permeability given by
the
difference in  the permittivity and the permeability  between  the exact and the approximate transformation media. Our formulation of cloaking as a
self-adjoint problem shows that the important issue of stability of cloaking  can be formulated as a problem in perturbation theory of the self-adjoint Maxwell
generator. Perturbation theory of self-adjoint operators is a main stream topic in modern mathematical physics and there is a large body of results.
See for example \cite{ka}. Our analysis opens the way to a rigorous study of the stability of cloaking along these lines.

For  exact transformation media the boundary conditions (\ref{4.1}, \ref{4.2}) are satisfied because they
follow from energy conservation and there is no need to add any lining to impose them. However, our
 results suggest a method to enhance
cloaking in the approximate transformation media that are used in
practice. Namely, to coat  $\partial  K$ by a material that imposes the boundary
conditions  (\ref{4.1}, \ref{4.2}). As these boundary conditions have to be satisfied
for  exact transformation media, adding a lining that enforces them
in the case of approximate transformation media will  improve the
performance of approximate cloaks.

\noindent {\bf Acknowledgement}

\noindent I thank Rainer Picard for his information on his work on the self-adjoint realizations of the
operator curl and of the Maxwell generator.

\vspace{10cm}
\begin{center}
\begin{figure}
\includegraphics[bb=0 490  650  790,clip,width=15cm]{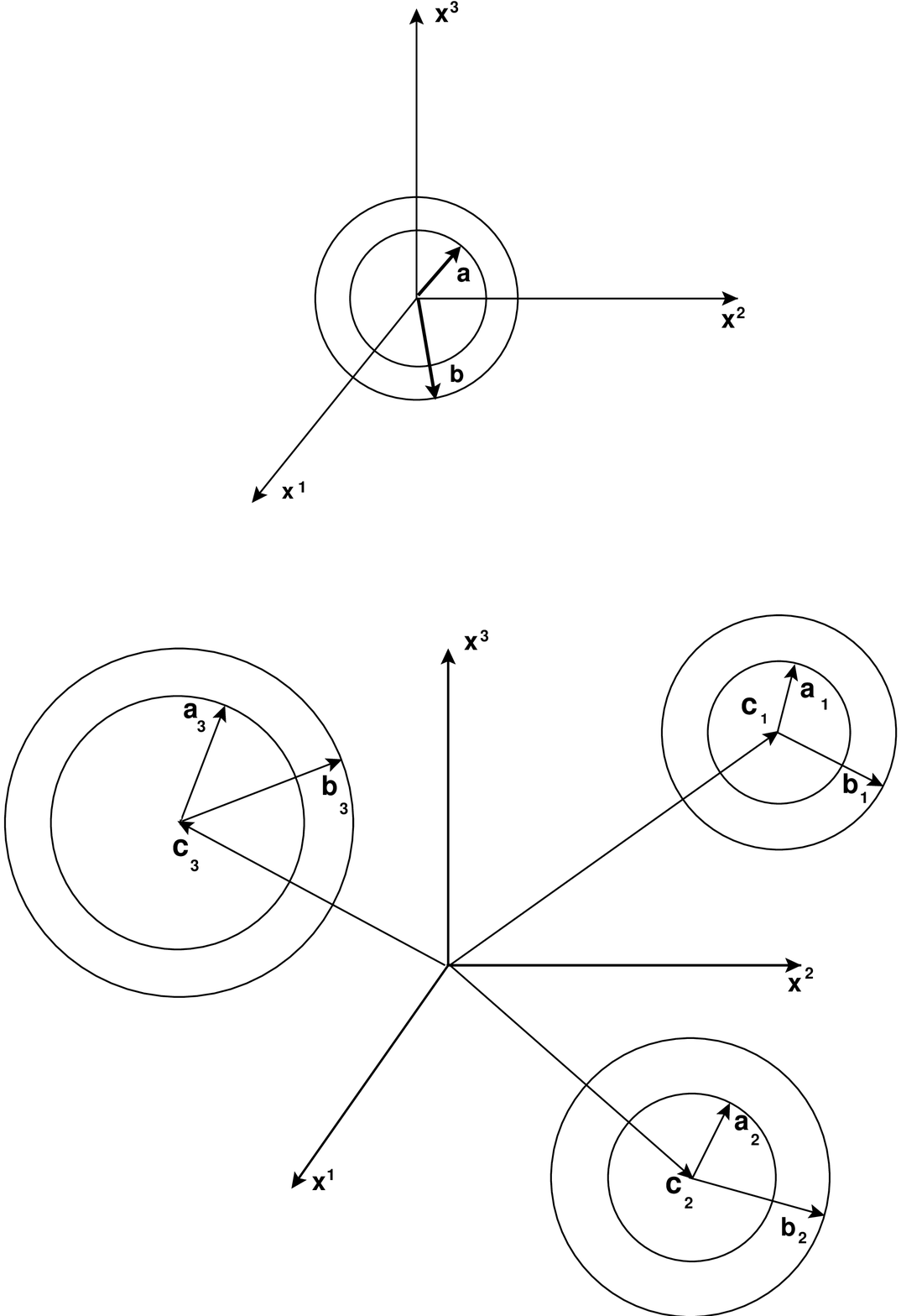}
\caption{One spherical cloak centered at zero.   }
\label{fig1}
\end{figure}

\begin{figure}
\includegraphics[bb=0 0  650  450,clip,width=15cm]{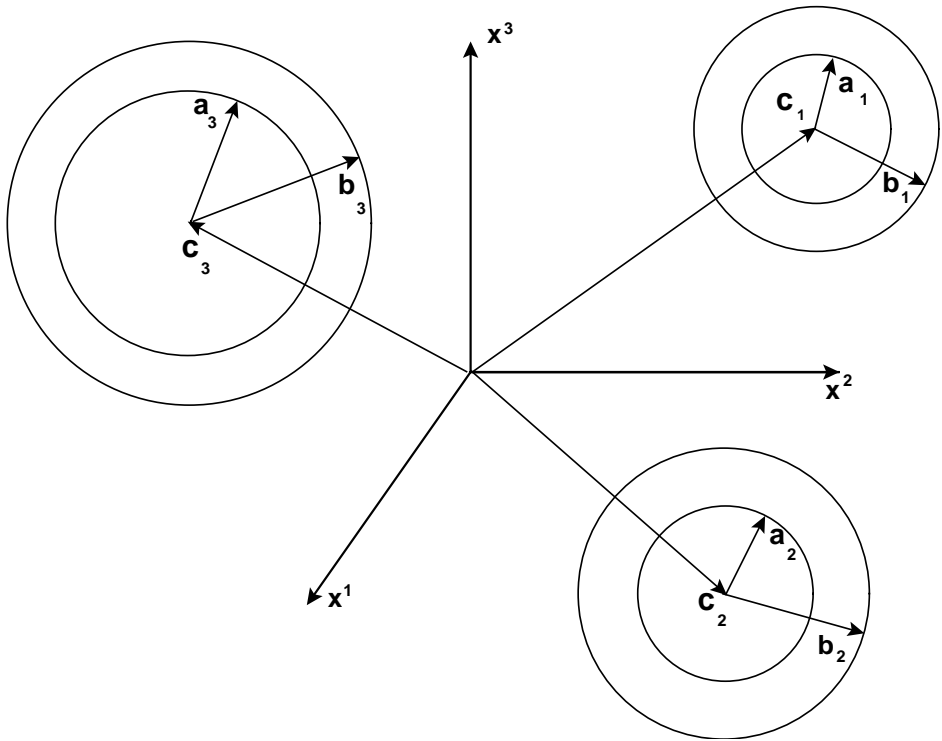}
\caption{Three spherical cloaks centered at $\c_1,\c_2,\c_3$.}
\label{fig2}
\end{figure}
\end{center}


\begin{thebibliography}{99}

\bibitem{pss1} J. B. Pendry, D. Schurig, D. R. Smith, Controlling electromagnetic fields,
Science {\bf 312} 1780-1782 (2006).

\bibitem{hen} A. Hendi, J. Hen, U. Leonhardt, Ambiguities in the
scattering tomography for central potentials, Phys. Rev. Lett. {\bf
97} 073902 (2006).

\bibitem{ho} W. Cai, U. K. Chettiar, A. V. Kildishev, G. W. Milton, V. M. Shalaev, Non-magnetic cloak without
reflection, arXiv: 0707.3641, 2007.

\bibitem{we1} R. Weder, A rigorous time-domain analysis of full-wave electromagnetic cloaking (invisibility),
arXiv: 0704.0248, 30 April  2007.

\bibitem{we2} R. Weder, A Rigorous analysis  of high order electromagnetic invisibility  cloaks,  J.  Phys A: Math.  Theor.
{\bf41} 065207 (2008).

\bibitem{yyrq} W. Yan, M. Yan, Z. Ruan, M. Qiu, Coordinate
transformations make perfect invisibility cloaks with arbitrary
shape, New J. Phys {\bf 10} 043040 (2008).

\bibitem{gklu} A. Greenleaf, Y. Kurylev, M. Lassas, G. Uhlmann,
Full-Wave invisibility of active devices at all frequencies, Commun.
Math. Phys. {\bf 275} 749-789 (2007).

\bibitem{vb} J. Van Bladel, Electromagnetic Fields, Hemisphere, Washington, 1985.

\bibitem{vhu} H. C. van de Hulst, Light Scattering by Small Particles, Dover, New York, 1957.



\bibitem{ccks} W. Cai, U.K. Chettiar, A.V. Kildishev,  V.M. Shalaev, Optical cloaking with metamaterials,
Nature Photonics {\bf 1} 224-226 (2007).

\bibitem{cpss} S.A. Cummer, B.-I. Popa, D. Schurig, D. R. Smith, J. Pendry, Full-wave simulation
of electromagnetic cloaking structures, Phys. Rev. E
{\bf 74} 036621 (2006).

\bibitem{smjcpss} D. Schurig,  J.J. Mock, B.J. Justice, S.A. Cummer, J.B. Pendry, A.F. Starr, D. R.
Smith, Metamaterial electromagnetic cloak
at microwave frequencies, Science {\bf 314} 977-980 (2006).

\bibitem{cwzk} H. Chen, B.-I. Wu, B. Zhang, J.A. Kong, Electromagnetic wave interactions with a metamaterial
cloak, Phys. Rev. Lett. {\bf 99} 063903 (2007).

\bibitem{zcwk} B. Zhang, H. Chen, B-I. Wu, J.A. Kong, Extraordinary surface voltage effect in the invisibility
cloak with an active device, Phys. Rev. Lett. {\bf 100} 063904
(2008).

\bibitem{le}  U. Leonhardt, Optical conformal mapping, Science {\bf 312} 1777-1780 (2006).

\bibitem{sps} D. Schurig, J.B. Pendry, D. R. Smith, Calculation of material properties and ray tracing in
transformation media, Opt. Exp. {\bf 14}
9794-9804 (2006).

\bibitem{shd}  I.I. Smolyaninov, Y.J.Hung, C.C. Davis, Electromagnetic cloaking in the visible frequency
range, arXiv:0709.2862, 2007.

\bibitem{gklu2} A. Greenleaf, Y. Kurylev, M. Lassas, G. Uhlmann, Improvement of cylindrical cloaking with the SHS lining,
 Optics Express {\bf 15} 12717-12734 (2007).


\bibitem{rynq} Z. Ruan, M. Yan, C. W. Neff, M. Qiu, Ideal Cylindrical Cloak: Perfect but Sensitive to
Tiny Perturbations, Phys. Rev. Lett. {\bf 99} 113903 (2007).

\bibitem{zcwrk} B. Zhang, H. Chen, B-I. Wu, Y. Luo, L. Ran, J.A.Kong, Response of a cylindrical invisibility
cloak to electromagnetic waves, Phys. Rev. B {\bf 76} 121101 (2007).



\bibitem{glu1} A. Greenleaf, M. Lassas, and G. Uhlmann, Anisotropic conductivities that cannot be detected by EIT, Physiol. Meas. {\bf 24}
413-419 (2003).

\bibitem{glu2} A. Greenleaf, M. Lassas, and G. Uhlmann, On nonuniqueness for Calder\'on's inverse problem, Math. Res. Let. {\bf 10} 685-693 (2003).







\bibitem{po} E.J. Post,  Formal Structure of Electromagnetics General Covariance and Electromagnetics,
Dover Publications, Mineola, New York, 1997.

\bibitem{pi1} R. Picard, Zur L\"osungstheorie der zeitunabh\"angiger Maxwellschen Gleichungen mit der
Randbedingung $ \mathbf n\cdot B= \mathbf n \cdot D =0$ in anisotropen, inhomogenen  Medien, Manuscripta
Math. {\bf 13} 37-52 (1974).


\bibitem{pi2} R. Picard, Ein Randwertproblem f\"ur die zeitunabh\"angigen Maxwellschen Gleichungen mit der Randbedingung
$n\cdot \varepsilon E=n\cdot \mu H=0$ in beschr\"anken Gebieten beliebigen Zusammenhangs, Appl. Anal. {\bf 6}
 207-221 (1977).

\bibitem{pi3} R. Picard, On a self-adjoint realization of curl and some of its applications, Ric. Mat.
{\bf XLVII} 153-180 (1998).

\bibitem{fi} N. Filonov, Spectral analysis of the selfadjoint operator curl in a region of finite measure,
St. Petersburg Math. J. {\bf 11} 1085-1095 (2000).



\bibitem{as} M. Abramowitz, I. A. Stegun, Handbook of Mathematical Functions, Dover, New York, 1970.

\bibitem{ka} T. Kato, Perturbation Theory of Linear Operators, Springer, Berlin, 1995.










\end{thebibliography}
\end{document}